# Revealing the spin optics of conics

Yanjun Bao[1], Shuai Zu[1], Wei Liu[1], Lei Zhou[2], Xing Zhu[1], and Zheyu Fang[1]*

[1]*School of Physics, State Key Lab for Mesoscopic Physics, Peking University, Beijing 100871, China*

[2]*State Key Laboratory of Surface Physics, Key Laboratory of Micro and Nano Photonic Structures (Ministry of Education) and Physics Department, Fudan University, Shanghai 200433, China*

e-mail address: zhyfang@pku.edu.cn

Ellipse and hyperbola are two well-known curves in mathematics with numerous applications in various fields, but their properties and inherent differences in spin optics are less understood. Here, we investigate the peculiar optical spin properties of the two curves and establish a connection between their foci and the spin states of incident light. We show that the optical spin Hall effect is the intrinsic optical spin property of ellipse, where photons with *different* spin states can be exactly separated to each of its two foci. While a hyperbola exhibits optical spin-selective effect, where only photons with one *particular* spin state can be accumulated at its foci. These properties are then experimentally demonstrated in near field by arranging nanoslits in conic shape. Based on the spin properties of the curves, we design spin-based plasmonic devices with various functionalities. Our results reveal the intrinsic optical spin properties behind conic curves and provide a route for designing spin-based plasmonic device.

# Introduction

Recently, a new branch of optical field, spin optics, has emerged and attracted considerable attention. By utilizing the spin angular momentum of light, this field concerns the spin-dependent optical phenomena, such as optical spin Hall effect [1-3], where the spin degeneracy of photons is removed through spin-orbit interaction, spin-dependent plasmonic focusing lens [4], spin-dependent optical vortex [5, 6], et al. On the other hand, metasurfaces, artificially subwavelength-structured interfaces, are able to control the phase of light, leading to versatile optical functionalities [7-10]. The use of metasurfaces in spin optics provides a flexible platform for applications in optical spin Hall effect [11, 12], optical Rashba effect [13] and holography [14-17].

Ellipse and hyperbola are two well-known conic curves with interesting properties in different physical fields. For example, in geometric optics, light emanating from (directing at) a focus of an elliptical (hyperbolic) mirror can be reflected to the other one. However, the properties of the two curves in spin optics are less known while spin optics and conic curves were both extensively studied within their own fields. It raises the question of what are the optical spin properties behind the two conic curves?

Here, for the first time, we investigate the optical spin properties of elliptical and hyperbolic curves and reveal the intrinsic difference between them. By introducing a geometric-phase distribution along the conic curves, we demonstrate that optical spin Hall effect and optical spin-selective effect are the intrinsic spin properties of ellipse and hyperbola, respectively. The foci of the two conic curves have a strong connection with both phenomena. To realize in practice, nanoslits as the unit elements are arranged on the elliptical and hyperbolic traces to provide the geometric phase. By carefully designing the positions and orientations of each nanoslit, the spin properties of the two conic curves are verified experimentally in the near field by using scanning near-field optical microscopy (SNOM). Furthermore, the spin optical properties of the conic curves are utilized to design spin-based plasmonic devices with various functionalities.

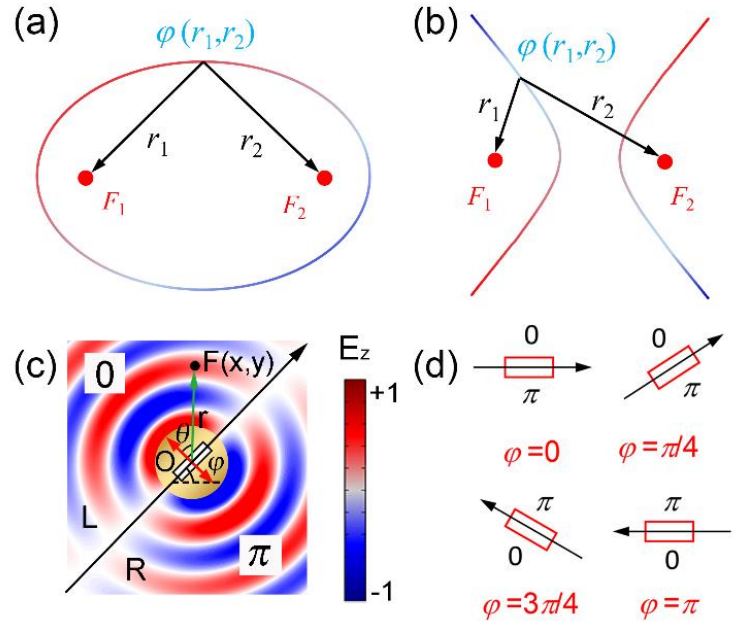

Fig. 1 (color online). (a-b) Schematic of ellipse (a) and hyperbola (b) with a geometric-phase distribution $\varphi$ ($r_1$, $r_2$) indicated with different colors along the conic traces. (c) Calculated normal component of the SPP electric field $E_z$ launched by a nanoslit perforated in an Au film. Red arrow: the polarization of incident light. $\varphi$: the azimuthal angle of nanoslit. $\theta$: the angle between the vector $OF$ and the short axis of nanoslit. (d) Anti-phase distributions $\psi$ of four nanoslits with different azimuthal angles.

## Results and Discussions

As shown in Fig. 1(a) and 1(b), the traces of ellipse and hyperbola are determined by $r_1 \pm r_2 = constant$, where $r_1$ and $r_2$ are the distances from any point at the traces to their two foci $F_1$ and $F_2$, respectively. To establish a connection with spin optics, we purposely rewrite the equations of ellipse and hyperbola curves in the following way

$(kr_1 \pm \varphi) + (kr_2 \mp \varphi) = constant$ (ellipse) (1)

$(kr_1 \pm \varphi) - (kr_2 \pm \varphi) = constant$ (hyperbola) (2)

where $k$ is the wave vector of light, $\sigma\varphi$ is the geometric phase and $\sigma = \pm 1$ corresponds to the incident spin state $|\sigma_\pm\rangle$. By introducing the geometric-phase term into the conic equations, we can link the two seemingly irrelevant fields, conics and spin optics. Take the ellipse as an example, if the introduced geometric-phase distribution satisfies that the phase term within the first bracket $kr_1 \pm \varphi$ is a constant value (modulo $2\pi$) along the elliptical curve, light with the spin state $|\sigma_\pm\rangle$ can constructively interfere at point $F_1$. At the same time, the phase term within the second bracket is automatically a constant value, which indicates that the light with opposite spin state constructively interferes at point $F_2$. This phenomenon can be considered as optical spin Hall effect, where photons with different spin states can be exactly separated to each of the two foci of ellipse. Similarly, there exists geometric-phase distribution that the phases within the two brackets in Eq. (2) can be both constant values along the hyperbolic curve. In this case, only photons with one particular spin state can be accumulated (constructively interfering) at both foci of hyperbola, as the optical spin-selective effect. The two optical spin effects are the intrinsic spin properties of the two conic curves. In the above two equations, the geometric phases in the two brackets are opposite for the ellipse but are the same for the hyperbola. It is this difference that leads to the different optical spin properties of the two curves.

Because the spin properties of conics are retrieved from their pure conic equations, their properties proposed here are general and can be demonstrated in different systems. In the following, we will demonstrate their properties in near field by using nanoslit, which can provide a suitable geometric phase. The properties of conics can also be demonstrated in other systems, such as in far field, as shown in our follow-up work [18]. Considering that nanoslit has been used to design structures for spin control [11, 19], we emphasize that the purpose of this work is to reveal the intrinsic optical spin properties behind these conic curves, which is totally different from the works that are aimed to design structures for optical spin Hall effect.

We firstly consider the surface plasmon polaritons (SPPs) excited by a nanoslit $O$ perforated in an Au film, as shown in Fig. 1(c). The black arrow indicates the direction of its long axis with an azimuthal angle $\varphi$, which is limited between 0 to $\pi$ (nanoslits oriented at $\varphi$ and $\varphi+\pi$ are equivalent). SPPs are preferentially generated when the normal incident light is polarized perpendicular to its long axis. The SPP emission is approximately that of an in-plane point dipole and exhibits $\pi$-phaseshift between the fields launched towards either side relative to the long axis of nanoslit. To account for the phase shift, we define an anti-phase term $\psi(O,F)$ is equal to 0 if a point $F$ is located at the left-hand-side of the black arrow and $\pi$ otherwise. Figure 1(d) shows this anti-phase distributions for several cases with different angles $\varphi$. Noticeably, the phase patterns are totally opposite for the two cases with $\varphi=0$ and $\pi$, although they are indeed the same. This contradiction can be explained by including the geometric phase, which will be introduced below.

When the structure was illuminated by circularly polarized (CP) light which is denoted with a spin state $|\sigma_\pm\rangle$, an additional geometric phase $\sigma\varphi$ is added due to the spin-orbit interaction [20, 21]. The $z$ component of SPP field at a point $F$ excited by the nanoslit $O$ can be written as [22]

$$E_{spp}(x,y) \propto \frac{\cos(\theta) e^{-\gamma r}}{\sqrt{r}} e^{i(k_{spp}r + \sigma\varphi + \psi)} \qquad (3)$$

where $r$ is the distance between $O$ and $F$, $k_{spp}$ and $\gamma$ are the real and image parts of the propagation constant of SPP, respectively, $\theta$ is the angle between the vector $OF$ and the short axis of nanoslit, $\sigma=\pm$

1 corresponds to the incident spin state $|\sigma_{\pm}\rangle$ and $\psi$ is the anti-phase term. By including the geometric phase, the phase term $\sigma\varphi+\psi$ are the same for the two nanoslits with $\varphi=0$ and $\pi$. Because the angle $\varphi$ can vary from 0 to $\pi$, and $\psi$ can be either 0 or $\pi$, $\sigma\varphi+\psi$ can cover the full range of $2\pi$ phase. However, for a fixed point $F$ and nanoslit, the range of this phase is limited. For example, when the nanoslit and point $F$ are both located at $x$ axis, and the azimuthal angle are varied from 0 to $\pi$, the point $F$ is always located at the same-hand-side of the nanoslit and thus $\sigma\varphi+\psi$ can only cover $\pi$ phase range.

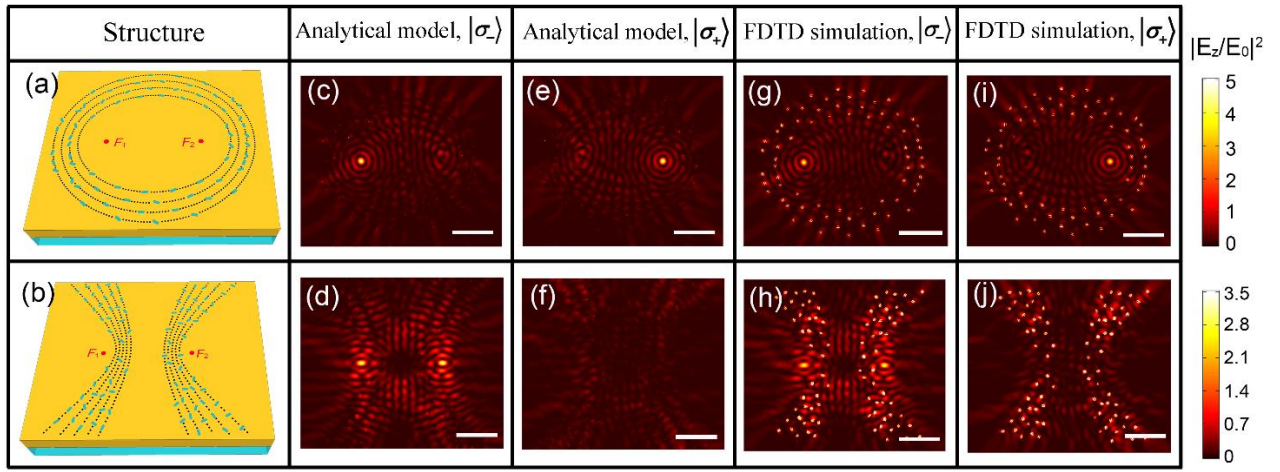

Fig. 2 (color online). (a-b) Schematics of an ellipse-shaped (a) and hyperbola-shaped (b) metasurfaces. (c-j) Analytical model (c-f) and FDTD simulation (10 nm distance above the metasurface) (g-j) calculated near-field intensity $|E_z|^2$ of the two conic-shaped metasurfaces with $|\sigma_{\pm}\rangle$ illuminations. The FDTD simulated intensity is normalized to incident electric intensity $|E_0|^2$. The numerical values shown on color bar apply only to the FDTD simulation intensity. The scale bars are 2 μm.

The geometric-phase distribution (the angle of nanoslit) along the conic curves can be determined from the constructive conditions. For example, for nanoslit at elliptical curve, it should fulfill the condition with $|\sigma_-\rangle$ spin state concentrating at point $F_1$: $-\varphi+\psi(O,F_1)+k_{spp}r_1=2\pi m+\varphi_1$, where $m$ is an integer and $\varphi_1$ is a constant value. At the same time, the geometric-phase distribution will automatically fulfill another constructive condition with $|\sigma_+\rangle$ spin state at point $F_2$: $\varphi+\psi(O,F_2)+k_{spp}r_2= 2\pi n+\varphi_2$, where $n$ is an integer and $\varphi_2$ is a constant value. Special care has to be taken for the anti-phase term and the limitation of the angle $\varphi$, which may cause the nonexistence of nanoslit at some positions of the elliptical curve. By adding the two equations above, we have $r_1+r_2= (p/2+\varphi_0)\lambda_{spp}$ [23], where $p$ is integer, $\varphi_0$ is a constant value and $\lambda_{spp}=2\pi/k_{spp}$ is the SPP wavelength. Thus the nanoslits can be located at a series of elliptical curves with common foci but with different major axes. Similarly, the geometric-phase distribution along the hyperbolic curve can be obtained from their constructive conditions [23].

Figure 2(a) shows our designed ellipse-shaped metasurface with four elliptical traces that are etched on a 100 nm-thickness Au film [23]. The hyperbola-shaped metasurface in Fig. 2(b) is designed for $|\sigma_-\rangle$ incidence with ten hyperbolic traces (It is also possible to design for $|\sigma_+\rangle$ incidence). The black dashed lines are drawn for the eye guide of each traces. Nanoslits at different positions possess different azimuthal angles (geometric phase) and are not fully distributed along the conic curves. This is because, on one hand, there is a restriction of a minimum gap of 300 nm between two adjacent nanoslits to avoid structural overlapping, and on the other hand the phase of nanoslit cannot cover the full range of $2\pi$ phase for a fixed point. Both structures were designed for operation at wavelength of $\lambda=671$ nm. Each nanoslit can be treated as a dipole with its SPP intensity distribution given by Eq. (3). The total field can be calculated analytically as the sum of the SPPs excited by each nanoslit. The analytical results for the two metasurfaces with $|\sigma_\pm\rangle$ incidences are shown in Fig. 2(c) -2(f). We also performed finite-difference time-domain (FDTD) simulations [23] and the simulated $|\mathbf{E}_z|^2$ fields are shown in Fig. 2(g)-2(j). It can be seen that the analytical model and FDTD simulations almost give the exact same results. For current structures, the intensities at the foci are increased by 5-fold and 3.5-fold for ellipse- and hyperbola-shaped metasurfaces, respectively. This can be further increased if more nanoslits are introduced or placed closer to the foci. The results from both methods show that the photons with different spin states are able to be separated to each of the elliptic foci, which is a clear demonstrating of optical spin Hall effect. While for hyperbola-shaped metasurface, only photons with $|\sigma_-\rangle$ spin state can be focused at its both foci, as a manifestation of optical spin-selective effect.

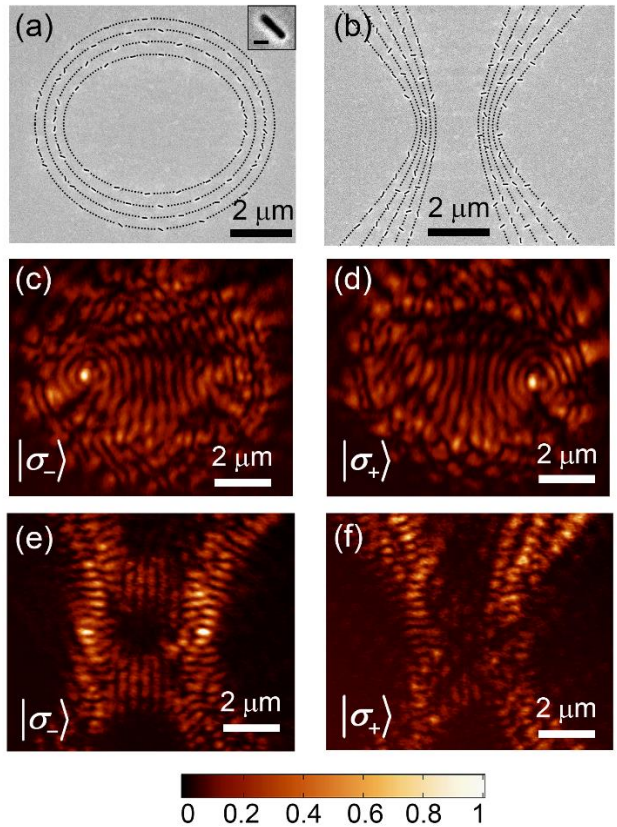

Fig. 3 (color online). (a-b) SEM images of the ellipse-shaped (a) and hyperbola-shaped (b) metasurfaces. The inset in (a) is a SEM image of a nanoslit with width of 50 nm and length of 200 nm (scale bar: 100 nm). The black dashed lines are a guide for the eye of conic traces. (c-f) Measured near-field optical intensity profiles of the ellipse-shaped (c-d) and hyperbola-shaped (e-f) metasurfaces with $|\sigma_\pm\rangle$ illuminations. The spin states of incident photons are indicated in each panel.

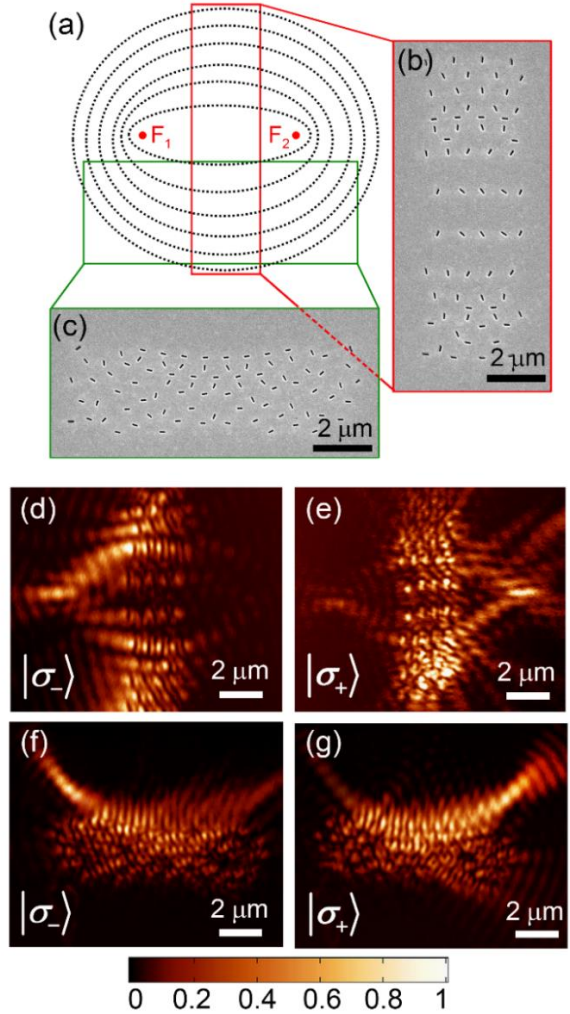

Fig. 4 (color online). (a) Design strategy of metasurfaces with different functionalities. (b-c) SEM images of two metasurfaces, which can separate photons with opposite spin states at each side of it (b) or at the same (upper) side (c). (d-g) Measured near-field optical intensities for metasurfaces shown in b (upper row) and c (lower row) with $|\sigma_\pm\rangle$ illuminations. The spin states of incident photons are indicated in each panel.

To experimentally demonstrate the concept, we fabricated the conic-shaped metasurfaces by using Au electron-beam evaporation and focused ion-beam milling [23]. Figure 3(a) and 3(b) present a scanning electron microscopy (SEM) image of the fabricated ellipse-shaped and hyperbola-shaped metasurfaces. A collection- mode SNOM system was used to measure SPP distributions [23]. A probe with a large half-cone angle ($\approx 15°$) is chosen and modified with sharp perturbation to increase the coupling efficiency of the **Ez** component [24]. The samples were back-illuminated with circular polarized light, which is generated by a quarter wave plate and a polarizer. Figure 3(c) and 3(d) show the measured near-field optical distribution of the ellipse-shaped metasurface, which clearly demonstrate that photons with $|\sigma_-\rangle$ spin state are focused at $F_1$, and switched to the right focus $F_2$ when the incident spin state is altered. For the hyperbola-shaped metasurface, only the photons with $|\sigma_-\rangle$ spin state can be accumulated at its both foci, as shown in Fig. 3(e) and 3(f). We note that there are some deviations between the simulation and experimental results, which may be arisen from the fabrication error, measurement accuracy and the affection of the in-plane electric field component.

As for practice applications, many structures with spin-dependent functionalities have been designed in the literature. For example, photons with different spin states can be delivered to different positions (or directions) that are located at the same side [25] or each side [19, 26] of a metasurface. In these structures, different functionalities require different design strategies. However, all these functionalities can be realized based on our conic-shaped metasurfaces without further design consideration. Take the ellipse-shaped metasurface for example, as shown in Fig. 4(a), points $F_1$ and $F_2$ are the common foci of the ellipses (dashed black lines) where nanoslits are located. By choosing parts of the elliptical traces, for example, regions between the two foci (red rectangle) or below (green rectangle), the resulting metasurface can focus photons with opposite spin states at each side of it or at the same (upper) side. We fabricated the two metasurfaces with their SEM images shown in Fig. 4(b) and 4(c). The measured SPP intensities with $|\sigma_\pm\rangle$ illuminations for the two metasurfaces are shown in Fig. 3(d)-3(g), which are in good agreements with our prediction and FDTD simulated results [23]. Actually, the positions of the two foci are not limited to the above two cases, but can be designed to locate at arbitrary positions relative to the metasurface [23]. Similarly, if hyperbolic curve is used, two foci can be simultaneously obtained at any two points relative to the metasurface with CP light. The conic-shaped metasurfaces, which are able to separate, select and deliver the incident light with spin polarization into any positions, provide a fast and flexible platform for designing spin-based devices.

In summary, we have theoretically and experimentally demonstrated the peculiar optical spin properties behind elliptical and hyperbolic curves. Optical spin Hall effect and optical spin-selective effect can be observed in ellipse-shaped and hyperbola-shaped metasurfaces with a strong connection with their foci, respectively. The properties of the two conic curves in geometric optics are known to everyone, and we hope that this work can give people a glimpse of their properties in spin optics, which may inspire the exploration of the properties of conic curves in other physical fields.

This work is supported by the National Basic Research Program of China (973 Program, Grant No. 2015CB932403), National Science Foundation of China (Grant No. 61422501, 11374023, 61521004, 11474057 and 11674068), Beijing Natural Science Foundation (Grant No. L140007), Shanghai Science and Technology Committee (16JC1403100), and Foundation for the Author of National Excellent Doctoral Dissertation of PR China (Grant No.201420), and National Program for Support of Top-notch Young Professionals.

# Supporting Information

**Contents:**

**1. The determination of the geometric-phase distribution (angle of nanoslit) along the conic curves.**

**2. The design of metasurfaces, sample fabrication and optical measurements.**

**3. The detail of the simulation procedure**

**4. Simulation results of the metasurfaces shown in Fig. 4**

**5. Design of metasurface for focusing photons at arbitrary positions**

**1. The determination of the geometric-phase distribution (angle of nanoslit) along the conic curves.**

For the ellipse, optical spin Hall effect requires that the angle of nanoslits should fulfill the following constructive conditions:

$$-\varphi + \psi(O, F_1) + 2\pi r_1/\lambda_{spp} = 2\pi n + \varphi_1 \tag{S1}$$

for $|\sigma_-\rangle$ illumination at point $F_1$, and

$$\varphi + \psi(O, F_2) + 2\pi r_2/\lambda_{spp} = 2\pi n + \varphi_2 \tag{S2}$$

for $|\sigma_+\rangle$ illumination at point $F_2$, where $m$ and $n$ are integers, $\varphi_1$ and $\varphi_2$ are constant values and $\lambda_{spp}$ is the SPP wavelength. The geometric-phase distribution can be obtained from either of the above two equations.

By adding the two equations above, we have

$$r_1 + r_2 = \left[ m + n - \frac{\psi(O, F_1) + \psi(O, F_2)}{2\pi} + \frac{\varphi_1 + \varphi_2}{2\pi} \right] \lambda_{spp} \quad (S3)$$

Because $\psi(O, F)$ is either 0 or $\pi$, the third term on the right side in the bracket of Eq. S3 is either 0, 0.5 or 1. So we can rewrite the above equation as follows:

$$r_1 + r_2 = (p/2 + \varphi_0) \lambda_{spp} \quad (S4)$$

where $p$ is an integer, and $\varphi_0 = \frac{\varphi_1 + \varphi_2}{2\pi}$.

For the hyperbola, optical spin selective effect require that only one particular spin state can be accumulated at point $F_1$ and point $F_2$ simultaneously. Take the photons with $|\sigma_-\rangle$ spin state for example, we have:

$$-\varphi + \psi(O, F_1) + 2\pi r_1 / \lambda_{spp} = 2\pi m + \varphi_1 \quad (S5)$$

at point $F_1$, and

$$-\varphi + \psi(O, F_2) + 2\pi r_2 / \lambda_{spp} = 2\pi n + \varphi_2 \quad (S6)$$

at point $F_2$. The geometric-phase distribution can be obtained from either of the above two equations.

By subtracting these two equations from each other, we get an equation that determines the coordinates of nanoslits as follows,

$$r_1 - r_2 = [p/2 + \varphi_0] \lambda_{spp} \quad (S7)$$

where $\varphi_0 = \frac{\varphi_1 - \varphi_2}{2\pi}$. This indicates that the nanoslits can be located at a series of hyperbolic curves with common foci but with different major axes.

**2. The design of metasurfaces, sample fabrication and optical measurements.**

For the metasurfaces shown in Figs. 2-3, the locations of foci are designed at $F_1$ (-2 μm, 0 μm) and $F_2$ (2 μm, 0 μm), respectively. For the metasurfaces shown in Fig. 4, the foci are $F_1$ (-5 μm, 0 μm) and $F_2$ (5 μm, 0 μm), respectively. The geometrical parameters of the nanoslit are identical with width of $w$ =50 nm and length of $L$=200 nm. The distance between two adjacent nanoslits has a minimum gap of 300 nm to avoid structural overlapping.

We evaporated 100 nm thickness Au film onto a glass substrate and used focused ion beam (FIB) milling to etch the nanoslit structures on the film. The evaporation rate was 0.5 Ås$^{-1}$ and the beam current was 1.1 pA to ensure the fabrication accuracy. A commercial SNOM system (Nanonics Imaging Multiview 1000) was used to measure the SPP distributions. The probe is metal-coated, tapered fiber probe with about half-cone angle of 15°. The probe was modified with sharp perturbation to increase the coupling efficiency of the **E**$_z$ component. In addition, because the **E**$_z$ component of SPPs is much larger than the in-plane **E**$_{x-y}$ component at the Au surface, the measured optical intensity is expected mostly come from the normal electric field |**E**$_z$|$^2$. The samples were back-illuminated with circular polarized light, which is generated by a quarter wave plate and a polarizer. Figure S1 presents a full schematic of the experimental setup.

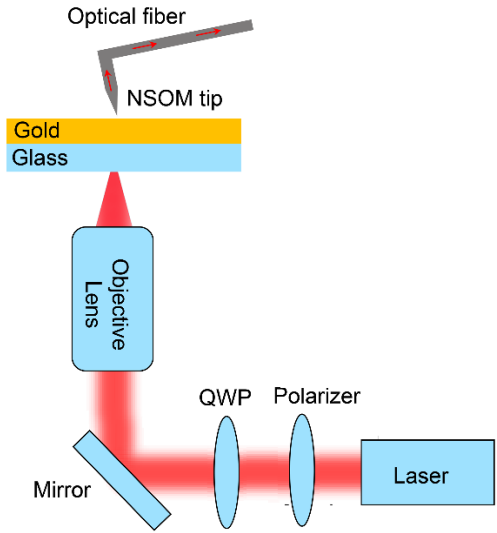

**Fig. S1** NSOM setup for measurement of the near field intensity.

### 3. The detail of the simulation procedure

Three dimensional vectorial electromagnetic simulations were performed by a commercial FDTD software (FDTD solutions, Lumerical). Total field scattered field (TFSF) was used as the incident source. The whole structure was surrounded by perfect match layers (PML) as the boundary condition. The dielectric constant of Au was taken from Johnson & Christy [1]. A plane monitor was placed 10 nm above the structure surface to detect the electric field intensity.

### 4. Simulation results of the metasurfaces shown in Fig. 4

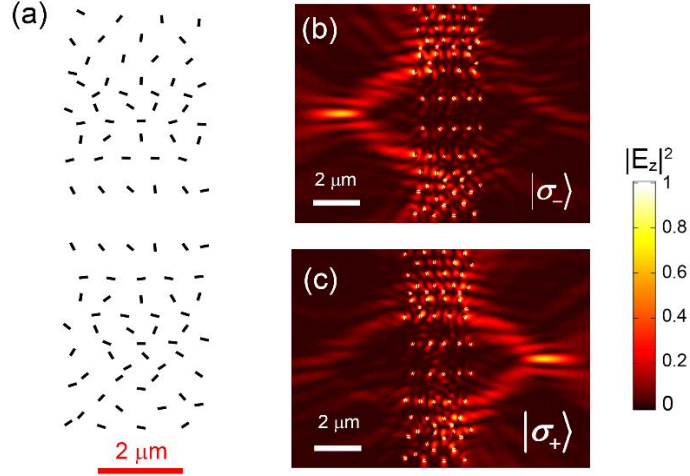

**Fig. S2** (a) The design pattern of the metasurface that can focus photons with opposite spin states at each side of it. (b-c) The simulated near-field intensities with $|\sigma_\pm\rangle$ illuminations.

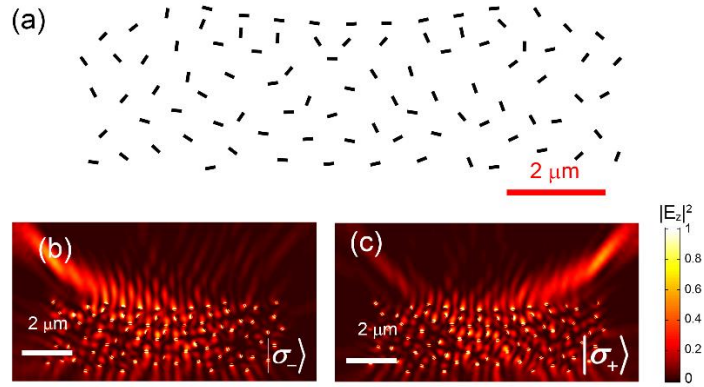

**Fig. S3** (a) The design pattern of the metasurface that can focus photons with opposite spin states at the same side of it. (b-c) The simulated near-field intensities with $|\sigma_\pm\rangle$ illuminations.

## 5. Design of metasurface for focusing photons at arbitrary positions

In Fig. 4 of the main text, we have only shown the two cases where the two foci are located at the same side and left-right side. However, the two foci can be located at arbitrary positions relative to the metasurfaces if the proper regions of ellipse-shaped metasurfaces are chosen. In Fig. S4, we have shown other four cases that the two foci can be located at right-down, up-left, up-right and left-down positions relative the square metasurfaces.

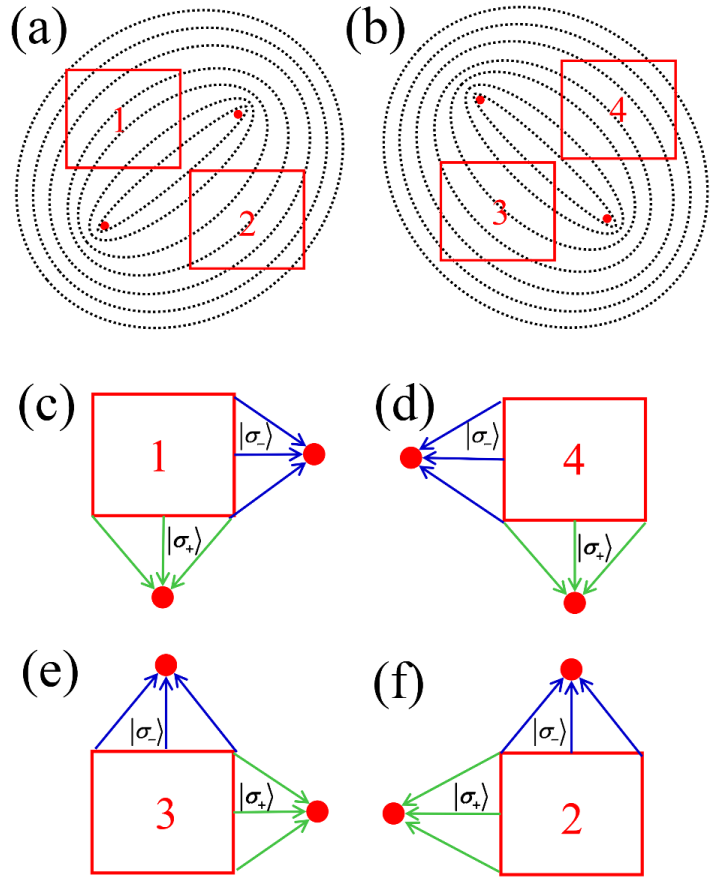

**Fig. S4** (a-b) The selecting strategies of the ellipse-shaped metasurface for delivering photons with different spin states at arbitrary positions. (c-f) Schematic of metasurfaces that can focus light with different spin states at right-down, left-down, up-right and up-left positions relative the square metasurfaces.